\documentclass[11pt]{cernrep}
\usepackage{graphicx}
\usepackage{here}
\usepackage{epsfig}
\include{rlFig}
\begin{document}
\def    \nn             {\nonumber}
\def    \=              {\;=\;}
\def    \frac           #1#2{{#1 \over #2}}
\def    \ret            {\\[\eqskip]}
\def    \ie             {{\em i.e.\/} }
\def    \eg             {{\em e.g.\/} }
\def    \lsim           {\raisebox{-3pt}{$\>\stackrel{<}{\scriptstyle\sim}\>$}}
\def    \gsim           {\raisebox{-3pt}{$\>\stackrel{>}{\scriptstyle\sim}\>$}}
\def    \gtrsim         {\raisebox{-3pt}{$\>\stackrel{>}{\scriptstyle\sim}\>$}}
\def    \esim           {\raisebox{-3pt}{$\>\stackrel{-}{\scriptstyle\sim}\>$}}
\newcommand     \be     {\begin{equation}}
\newcommand     \ee     {\end{equation}}
\newcommand     \ba     {\begin{eqnarray}}
\newcommand     \ea     {\end{eqnarray}}
\newcommand     \sst            {\scriptstyle}
\newcommand     \sss            {\scriptscriptstyle}
\newcommand     \avg[1]         {\left\langle #1 \right\rangle}
\newcommand     \Ca             {{C_{\rm A}}}
\newcommand     \Cf             {{C_{\rm f}}}
\newcommand     \lambdamsb     {\ifmmode
          \Lambda_4^{\rm \scriptscriptstyle \overline{MS}} \else
         $\Lambda_4^{\rm \scriptscriptstyle \overline{MS}}$ \fi}
\newcommand     \MSB            {\ifmmode {\overline{\rm MS}} \else
                                 $\overline{\rm MS}$  \fi}
\newcommand     \nf             {n_{\rm f}}
\newcommand     \nlf            {n_{\rm lf}}
\newcommand     \ptmin     {\ifmmode p_{\scriptscriptstyle T}^{\sss min} \else
                           $p_{\scriptscriptstyle T}^{\sss min}$ \fi}
\def     \muf           {\mbox{$\mu_{\sss F}$}}
\def     \mur            {\mbox{$\mu_{\sss R}$}}
\def    \muo            {\mbox{$\mu_0$}}
\newcommand\as{\alpha_{\sss S}}
\newcommand\astwo{\alpha_{\sss S}^2}
\newcommand\asthree{\alpha_{\sss S}^3}
\newcommand\asfour{\alpha_{\sss S}^4}
\newcommand\epb{\overline{\epsilon}}
\newcommand\aem{\alpha_{\rm em}}
\newcommand\QQb{{Q\overline{Q}}}
\newcommand\qqb{{q\overline{q}}}
\newcommand\cb{\overline{c}}
\newcommand\bb{\overline{b}}
\newcommand\tb{\overline{t}}
\newcommand\Qb{\overline{Q}}
\newcommand\qq{{\scriptscriptstyle Q\overline{Q}}}
\def \asopi{\mbox{$\frac{\as}{\pi}$}}
\def \oacube {\mbox{${\cal O}(\asthree)$}}
\def \oatwo {\mbox{${\cal O}(\astwo)$}}
\def \oas   {\mbox{${\cal O}(\as)$}}
\def \ppbar {\mbox{$p \bar p$}}
\def \ttbar {\mbox{$t \bar t$}}
\def \bbbar {\mbox{$b \bar b$}}
\def \ccbar {\mbox{$c \bar c$}}
\def \mtt   {\mbox{$M_{\scriptscriptstyle t\bar t}$}}
\def \pt   {\mbox{$p_{\scriptscriptstyle T}$}}
\def \ptpair   {\mbox{$p_{\scriptscriptstyle T}^{\scriptscriptstyle
                t\bar t}$}}
\def \et   {\mbox{$E_{\scriptscriptstyle T}$}}
\def \etsq {\mbox{$E_{\scriptscriptstyle T}^2$}}
\def \rap   {\mbox{$\eta$}}
\def \deltar {\mbox{$\Delta R$}}
\def \dphi {\mbox{$\Delta \phi$}}
\def \to   {\mbox{$\rightarrow$}}
\def    \mb             {\mbox{$m_b$}}
\def    \mc             {\mbox{$m_c$}}
\def    \mt             {\mbox{$m_t$}}
\newcommand \jpsi{\ifmmode{J/\psi
    }\else{$J/\psi$}\fi}
\def\calF{{\cal F}}
\def\calP{{\cal P}}
\def\calM{{\cal M}}
\def\calO{{\cal O}}
\newcommand{\Tr}{{\mbox{\rm Tr}}}
\newcommand{\mn}{{\mu\nu}}
\newcommand{\half}{{1\over 2}}
\newcommand{\bea}{\begin{eqnarray}}
\newcommand{\eea}{\end{eqnarray}}

\title{W$^+$W$^+$ Scattering as a Sensitive Test of the 
Anomalous Gauge Couplings of the Higgs Boson at the LHC
  \footnote{~~Contributed to Workshop on Physics at TeV Colliders, Les Houches, 
 France, 26 May -- 6 June 2003.}
                        }
\author{
H.-J. He$^1$, ~Y.-P. Kuang$^2$, ~C.-P. Yuan$^3$, 
~and ~B. Zhang$^2$
}

\institute{
$^1$Center for Particle Physics, University of Texas at Austin, Austin, 
Texas 78712, USA. \\
$^2$Center of High Energy Physics, Tsinghua University, 
Beijing 100084, China. \\
$^3$Department of Physics and Astronomy, Michigan State University,
East Lansing, MI 48824, USA.}

\maketitle

\begin{abstract}
We propose a sensitive way to test the anomalous $HVV$ couplings
($~V=W^\pm,~Z^0$) of the Higgs boson ($H$), which can arise from
either the dimension-3 effective operator in a nonlinearly
realized Higgs sector or the dimension-6 effective operators in a
linearly realized Higgs sector, via studying the VV scattering
processes at the CERN Large Hadron Collider (LHC). We show that, with
an integrated luminosity of 300 fb$^{-1}$ and sufficient kinematical 
cuts for suppressing the backgrounds, studying the process 
$pp \to W^+ W^+ j j \to \l^+ \nu \l^+ \nu j j$ 
can probe the anomalous HWW couplings at a few tens of percent level 
for the nonlinearly realized Higgs sector, and at the level of 
0.01-0.08 TeV$^{-1}$ for the linearly realized effective Lagrangian.

\end{abstract}

\setcounter{footnote}{0}
\renewcommand{\thefootnote}{\arabic{footnote}}

The electroweak symmetry breaking  mechanism (EWSBM) is one of the most
profound puzzles in particle physics. Since the Higgs
sector of the standard model (SM) suffers the well-known problems of triviality
and unnaturalness, there has to be new
physics beyond the SM above certain high energy scale $\Lambda$.
If a light Higgs boson candidate ($H$)
is found in future collider experiments,
the next important task is to experimentally measure the gauge
interactions of this Higgs scalar and explore the nature of the
EWSBM. Let $~V=W^\pm,~Z^0~$ be the electroweak (EW) gauge bosons.
The detection of the anomalous $HVV$ couplings (AHVVC) will point to new physics beyond the SM
underlying the EWSBM.
\begin{figure}[h]
\begin{center}
\includegraphics[width=0.75\textwidth,clip]{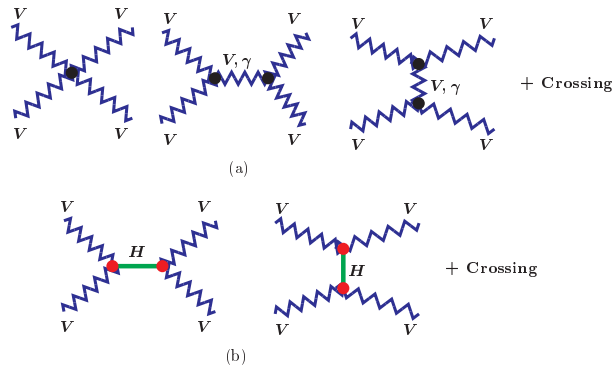}
\end{center}
\vskip -0.5cm
\caption{Illustration of Feynman diagrams for $VV$
scatterings in the SM: (a) diagrams contributing to $T(V,\gamma)$,
(b) diagrams contributing to $T(H)$.} \label{Fig.1}
\end{figure}

Before knowing the correct new physics, the effect of new physics
at energy below $\Lambda$ can be parametrized as
effective operators in an effective theory.
This is a model-independent
description.
Testing the AHVVC relative to that of the SM can
discriminate the EWSBM in the new physics model
from that of the SM. In Ref. \cite{HKYZ03}, we propose
a sensitive way of testing the
AHVVC via $VV$ scatterings, especially the $W^+W^+$
scatterings, at the LHC \cite{HKYZ03}. This includes the test of
either the dim-3 AHVVC in a nonlinearly realized
Higgs model (NRHM) \cite{CK} or the dim-6 AHVVC
in the linearly realized effective interactions (LREI) \cite{linear}.
The reason for the sensitiveness is the following.
 The scattering amplitude
contains two parts: (i) the amplitude $T(V,\gamma)$ related only
to $V$ and $\gamma$ (Fig.\,1(a)), and (ii)
the amplitude $T(H)$ related to the Higgs boson (Fig.\,1(b)).
At high energies, both $T(V,\gamma)$ and $T(H)$
increase with the center-of-mass energy ($E$) as
$E^2$ in the NRHM and as $E^4$ in the LREI.
In the SM, though individual diagrams in Fig.\,1(a) may behave as
$E^4$, the sum of all diagrams in Fig.\,1(a) can have
at most $E^2$-dependent contribution.
The $HVV$ coupling constant
in the SM is just the non-Abelian gauge coupling constant. This causes the two
$E^2$-dependent pieces to precisely cancel with each other in
$T(V,\gamma)+T(H)$, resulting in the expected $E^0$-behavior for
the total amplitude, as required by the unitarity of the
$S$ matrix. If there is AHVVC due to new physics effect, $T(V,\gamma)+T(H)$
can grow as $E^2$ or $E^4$ in the high energy regime.
Such deviations from the $E^0$ behavior of the SM
amplitude can provide a rather sensitive test of the AHVVC in high energy
$VV$ scattering experiments.  This type
of tests do not require
the measurement of the $H$ decay branching
ratios, and is thus of special interest, especially if the AHVVC
are very large or very small \cite{HKYZ03}.

We take such enhanced $VV$ scatterings as the signals for
testing the AHVVC. To avoid the large hadronic
backgrounds at the LHC \cite{Chanowitz}, We choose the gold-plated pure leptonic
decay modes of the final state $V$s as the tagging modes.
Even so, there are still several kinds of backgrounds to be
eliminated \cite{WW94,WW95}.
We take all the kinematic cuts given in Ref. \cite{WW95} to suppress the
backgrounds, and calculate the complete tree level
contributions to the process
\begin{eqnarray}                        
pp\to VVjj\to llll(\nu\nu)jj,
\label{VVjj}
\end{eqnarray}
where $j$ is the forward jet that is tagged
to suppress the large background rates.
Our calculation shows that, for
not too small AHVVC, all the
backgrounds can be reasonably suppressed by such kinematic cuts.
In the case of the SM, there are still considerably large remaining backgrounds
contributed by the transverse component $V_T$. We
shall call these the {\it remaining SM backgrounds}
(RSMB) after taking the above treatment. Our calculation shows
that the signals can be considerably larger than the RSMB even
with not very large AHVVC.

We first consider the NRHM.
The effective Lagrangian below $\Lambda$, up to dim-4 operators,
respecting the EW gauge symmetry,
charge conjugation, 
parity,
and the custodial $SU(2)_c$ symmetry, is \cite{CK}:.
\begin{eqnarray}                            
{\cal L}&=&-\frac{1}{4}{\overrightarrow
W}_{\mu\nu}\cdot{\overrightarrow
W}^{\mu\nu}-\frac{1}{4}B_{\mu\nu}B^{\mu\nu}
+\frac{1}{4}(v^2+2\kappa vH+\kappa^\prime H^2){\rm
Tr}(D_\mu\Sigma^\dagger D^\mu\Sigma)\nonumber\\
&&+\frac{1}{2}\partial_\mu H\partial^\mu H
-\frac{m_H^2}{2}H^2-\frac{\lambda_3 v}{3!}H^3+\frac{\lambda_4}{4!}H^4,
\label{Lagrangian}
\end{eqnarray}
where $\overrightarrow W_{\mu\nu}$ and $B_{\mu\nu}$ are field
strengths of the EW gauge fields, $v\simeq 246$\,GeV is the
vacuum expectation value (VEV) breaking the EW gauge symmetry,
$(\kappa,\,\lambda_3)$ and
$(~\kappa^\prime,\,\lambda_4)$ are, respectively,
dimensionless coupling constants from the dim-3 and
dim-4 operators,
$\Sigma=\exp\{i{\overrightarrow\tau}\cdot{\overrightarrow\omega}/
{v}\}$,
and $~D_\mu\Sigma=
\partial_\mu\Sigma+ig\frac{\overrightarrow\tau}{2}\cdot{\overrightarrow
W}_\mu\Sigma -ig'B_\mu\Sigma\frac{\tau_3}{2}$.
The SM corresponds to
$\kappa=\kappa^\prime=1$ and
$\displaystyle\lambda_3=\lambda_4 =\lambda={3 m_H^2}/{v^2}$.

At the tree level, only the dim-3 operator
$\frac{1}{2}\kappa vHD_\mu\Sigma^\dagger D^\mu\Sigma$
contributes to the $VV$ scatterings in Fig.\,1. Therefore,
$VV$ scatterings can test
$\kappa$, and $\Delta\kappa\equiv\kappa -1$ measures the deviation from the
SM value $\kappa =1$.

In Ref. \cite{HKYZ03}, the full tree level cross sections for all the processes
in (\ref{VVjj}) are calcuated for $115~{\rm GeV}\le m_H\le 300$ GeV.
The results show that the most sensitive channel
is ~$pp\to W^+W^+jj\to l^+\nu l^+\nu jj$ \cite{HKYZ03}. With an integrated
luminosity of 300 fb$^{-1}$, there are more than 20 events for
$\Delta\kappa\ge 0.2$ or $\Delta\kappa\le -0.3$, while there are only
about 15 RSMB events (see Ref. \cite{HKYZ03} for details).
Considering only the statistical errors,
the LHC can limit $\Delta\kappa$ to the range
\begin{eqnarray}                    
-0.3<\Delta\kappa<0.2
\label{constraint}
\end{eqnarray}
at roughly the $(1-3)\sigma$ level if data is consistent with the
SM prediction \cite{HKYZ03}.

Other constraints on $\Delta\kappa$ from the precision EW data,
the requirement of the unitarity of the $S$-matrix, etc. were
studied in Ref.~\cite{HKYZ03}, which are either weaker than Eq.
(\ref{constraint}) or of the similar level \cite{HKYZ03}.

Next, we consider the LREI. In this theory, the leading AHVVC are from the effective
operators of dim-6 \cite{linear,G-G}.
As is shown in Refs.~\cite{linear,G-G}, the {\tt C} and {\tt P}
conserving effective Lagrangian up to dimension-6 operators
containing a Higgs doublet $\Phi$ and the weak bosons $V^a$ is
given by
\begin{equation}                    
{\cal L}_{\mbox{eff}} ~\,=~\, \sum_n \frac{f_n}{\Lambda^2} {\cal O}_n \,,
\label{l:eff}
\end{equation}
where ${\cal O}_n$'s are dim-6 operators composed of $\Phi$ and the EW gauge
fields (cf. Ref. \cite{G-G}), $f_n/\Lambda^2$'s are the AHVVC.

The precision EW data and the requirement of the unitarity of the $S$-matrix
give certain constraints on the $f_n$'s. The constraints on
$f_{WWW}/\Lambda^2$, $f_{WW}/\Lambda^2$, $f_{BB}/\Lambda^2$,
$f_W/\Lambda^2$, and $f_B/\Lambda^2$ from the presently available
experimental data are rather weak \cite{HKYZ03}.
A better test of them is to study the $VV$ scatterings.
In ${\cal L}_{\mbox{eff}}$, the operator ${\cal
O}_{WWW}$ contributes to the triple and quartic $V$ boson
self-interactions which may not be directly related to the EWSBM,
we assume $f_{WWW}/\Lambda^2$ is small in the analysis.
and concentrate on the test of $f_{WW}/\Lambda^2$,
$f_{BB}/\Lambda^2$, $f_W/\Lambda^2$,
and $f_B/\Lambda^2$ . They are related to the following AHVVC
in terms of $H$, $W^\pm$, $Z$, and $\gamma$ \cite{G-G}:
\begin{eqnarray}                         
{\cal L}^H_{\rm eff}&=&g_{H\gamma\gamma}HA_{\mu\nu}A^{\mu\nu}
+g^{(1)}_{HZ\gamma}A_{\mu\nu}Z^\mu\partial^\nu H
+g^{(2)}_{HZ\gamma}HA_{\mu\nu}Z^{\mu\nu}
+g^{(1)}_{HZZ}Z_{\mu\nu}Z^\mu
\partial^\nu H\nonumber\\
&&
+g^{(2)}_{HZZ}HZ_{\mu\nu}Z^{\mu\nu}
+g^{(1)}_{HWW}(W^+_{\mu\nu} W^{-\mu}\partial^\nu H+{\rm h.c.})
+g^{(2)}_{HWW}HW^+_{\mu\nu}W^{-\mu\nu},
\label{LHeff}
\end{eqnarray}
where
\begin{eqnarray}                           
\displaystyle
&&g^{}_{H\gamma\gamma}=-\bigg(\frac{gm_W}{\Lambda^2}\bigg)\frac{s^2(f_{BB}
+f_{WW})}{2},
~~g^{(1)}_{HZ\gamma}=\bigg(\frac{gm_W}{\Lambda^2}\bigg)\frac{s(f_W-f_B)}{2c},
\nonumber\\
&&g^{(2)}_{HZ\gamma}=\bigg(\frac{gm_W}{\Lambda^2}\bigg)\frac{s[s^2f_{BB}
-c^2f_{WW}]}{c},
~~g^{(1)}_{HZZ}=\bigg(\frac{gm_W}{\Lambda^2}\bigg)\frac{c^2f_W+s^2f_B}{2c^2},
\nonumber\\
&&g^{(2)}_{HZZ}=-\bigg(\frac{gm_W}{\Lambda^2}\bigg)\frac{s^4f_{BB}
+c^4f_{WW}}{2c^2},
~g^{(1)}_{HWW}=\bigg(\frac{gm_W}{\Lambda^2}\bigg)\frac{f_W}{2},
~g^{(2)}_{HWW}=-\bigg(\frac{gm_W}{\Lambda^2}\bigg)f_{WW},\,\,\,\,\,
\label{g}
\end{eqnarray}
with $s\equiv \sin\theta_W$ and $c\equiv \cos\theta_W$.

The test of these AHVVC via $VV$ scatterings is quite different from that of
$\Delta\kappa$. The relevant operators ${\cal O}_n$'s contain two derivatives.
so, at high energies, the interaction vertices themselves behave
as $E^2$, and thus
the longitudinal $VV$ scattering amplitudes, $V_LV_L \to V_LV_L$,
grows as $E^4$, and those containing $V_T$
grow as $E^2$. Hence the scattering processes containing
$V_T$ actually behave as {\it signals} rather than backgrounds.

It is shown in Ref. \cite{HKYZ03} that the most sensitive channel is still
$pp\to W^+W^+jj\to l^+\nu l^+\nu jj$. Detailed calculations show that the
contributions of $f_B$ and $f_{BB}$
are small even if they are of the same order of magnitude
as $f_W$ and $f_{WW}$ \cite{HKYZ03}.
Hence, we take account of only $f_W/\Lambda^2$ and
$f_{WW}/\Lambda^2$ in the analysis.
If they are of the same order of magnitude,
the interference between them may be significant, depending on
their relative phase which undoubtedly complicates the
analysis. Hence, we perform a single parameter analysis,
i.e., assuming only one of them is dominant at a time.
In the case that $f_W$ dominates, the obtained numbers of events
in $pp\to W^+W^+jj$ $\to l^+\nu l^+\nu jj$
with an integrated luminosity of $300$ fb$^{-1}$ are more than 20
for $f_W/\Lambda^2\ge 0.85$ TeV$^{-2}$ or $f_W/\Lambda^2\le -1.0$ TeV$^{-2}$
, and the number of the RSMB events are still around 15 (see
Ref. \cite{HKYZ03} for details). If no AHVVC
effect is found at the LHC, we can set the following bounds on $f_W/\Lambda^2$
(in units of TeV$^{-2}$) when taking into account of only the statistical error:
\begin{eqnarray}                       
1\sigma:~~-1.0< f_W/\Lambda^2< 0.85,~~~~~~~~~~~
2\sigma:~~-1.4< f_W/\Lambda^2\leq 1.2. \,\,\,\,\,\label{fW}
\end{eqnarray}
In the case that $f_{WW}$ dominates,
the corresponding bounds are (in
units of TeV$^{-2}$):
\begin{eqnarray}                       
1\sigma:~~-1.6\leq f_{WW}/\Lambda^2<1.6,
~~~~~~~~~~2\sigma:~~-2.2\leq f_{WW}/\Lambda^2< 2.2. \label{fWW}
\end{eqnarray}
These are somewhat weaker than those in Eq.~(\ref{fW}).
From  Eqs. (\ref{fW}) and (\ref{fWW})
we obtain the corresponding bounds on $g^{(i)}_{HVV},~i=1,2$ (in units of TeV$^{-1}$):
\begin{eqnarray}                           
&&1\sigma:\nonumber\\
&&\hspace{0.2cm}-0.026< g^{(1)}_{HWW}< 0.022,
~~~-0.026< g^{(1)}_{HZZ}< 0.022,
~~~-0.014< g^{(1)}_{HZ\gamma}< 0.012,\nonumber\\
&&\hspace{0.2cm}
-0.083\leq g^{(2)}_{HWW}< 0.083,
~~~~~~~~0.032\leq g^{(2)}_{HZZ}< 0.032,
~~~-0.018\leq g^{(2)}_{HZ\gamma}< 0.018,
\label{1sigmabounds}
\end{eqnarray}
\begin{eqnarray}                     
&&2\sigma:\nonumber\\
&&\hspace{0.4cm}-0.036< g^{(1)}_{HWW}\leq 0.031,
~~~~~~~0.036< g^{(1)}_{HZZ}\leq 0.031,
~~~~~~~0.020< g^{(1)}_{HZ\gamma}\leq 0.017,\nonumber\\
&&\hspace{0.4cm}-0.11~\leq g^{(2)}_{HWW}< ~0.11,
~~~~-0.044\leq g^{(2)}_{HZZ}< 0.044,
~~~-0.024\leq g^{(2)}_{HZ\gamma}< 0.024.
\label{2sigmabounds}
\end{eqnarray}
These bounds are to be compared with the $1\sigma$ bound on
$g^{(2)}_{HWW}$ obtained from studying the on-shell Higgs boson production
via weak boson fusion at the LHC given in
Ref.~\cite{PRZ}, where $g^{(2)}_{HWW}$ is
parametrized as $g^{(2)}_{HWW}$$=1/\Lambda_{5}=g^2
v/\Lambda^2_{6}$. The obtained $1\sigma$ bound on $\Lambda_6$
for an integrated luminosity of 100 fb$^{-1}$ is about $\Lambda_6
\ge$ 1 TeV \cite{PRZ}, which corresponds to $g^{(2)}_{HWW}=1/\Lambda_5\le
0.1~ {\rm TeV}^{-1}$. We see that the $1\sigma$ bounds in
Eq. (\ref{1sigmabounds}) are all stronger than the above bound
given in Ref.~\cite{PRZ}. For an
integrated luminosity of 300 fb$^{-1}$, the bound
in Ref.~\cite{PRZ} corresponds roughly to a 1.7$\sigma$ level accuracy.
Comparing it with the results in Eq. (\ref{2sigmabounds}), we conclude that our 2$\sigma$ bound on
$g^{(2)}_{HWW}$ is at about the same level of accuracy,
while our 2$\sigma$ bounds on
the other five $g^{(i)}_{HVV}~(i=1,2)$ are all stronger than
those given in Ref.~\cite{PRZ}.

It has been shown in Ref.~\cite{HZZ} that the anomalous $HZZ$
coupling constants $g^{(1)}_{HZZ}$ and $g^{(2)}_{HZZ}$ can be
tested rather sensitively at the Linear Collider (LC) via the
Higgs-strahlung process $e^+e^-\to Z^*\to Z+H$ with $Z\to
f\bar{f}$. The obtained limits are $g^{(1)}_{HZZ}\sim
g^{(2)}_{HZZ}\sim O(10^{-3}\--10^{-2})$ TeV$^{-1}$ \cite{HZZ}.
Although the bounds shown in Eqs. (\ref{1sigmabounds}) and
(\ref{2sigmabounds}) are weaker than these LC bounds, $W^+W^+$
scattering at the LHC can provide the bounds on
$g^{(i)}_{HWW},~i=1,2$ which are not easily accessible at the LC.
So that the two experiments are complementary to each other.

Further discrimination of the effect of the AHVVC
from that of a strongly interacting EW symmetry breaking sector
with no light resonance will eventually demand a multichannel
analysis at the LHC by searching for the light Higgs resonance
through all possible on-shell production channels including
gluon-gluon fusion. Once the light Higgs resonance is confirmed,
$VV$ scatterings, especially the $W^+W^+$ channel, can provide
rather sensitive tests of various AHVVC for
probing the EWSB mechanism. So $VV$ scatterings are not only
important for probing the strongly interacting EWSBM when there is
no light Higgs boson, but can also provide sensitive test of the
AHVVC (especially the anomalous $HWW$
couplings) at the LHC for discriminating new physics from the SM
when there is a light Higgs boson.

\end{document}